\documentclass{article}
\usepackage{amssymb}

\usepackage{graphicx}
\usepackage{amsmath}

\input{tcilatex}

\begin{document}

\title{Entanglement and Dynamic Stability of Nash Equilibria in a Symmetric Quantum
Game.}
\author{A. Iqbal and A.H. Toor \\
Department of Electronics, Quaid-i-Azam University, \\
Islamabad, Pakistan.\\
email: qubit@isb.paknet.com.pk}
\maketitle

\begin{abstract}
We study the evolutionary stability of\ Nash equilibria (NE) in a symmetric
quantum game played by the recently proposed scheme of applying `identity'
and `Pauli spin flip' operators on the initial state with classical
probabilities. We show that in this symmetric game dynamic stability of a NE
can be changed when the game changes its form, for example, from classical
to quantum. It happens even when the NE remains intact in both forms.
\end{abstract}

\section{Introduction}

In a recent paper Eisert et al \cite{eisert} have shown that when the
players in a game have access to quantum strategies the game can give
different Nash Equilibria (NE) that can give better payoffs. Marinatto and
Weber\cite{marinatto} have proposed another interesting scheme to play a
quantum game where players implement their `tactics' on an initial strategy
by a probabilistic choice between applying the `identity' and `Pauli
spin-flip' operators. In a quantized game there can rise multiple NE similar
to a classical game. For example classical version of the Battle of Sexes
game gives three NE and its quantization by Marinatto and Weber's scheme
again gives three NE. In classical game theory dynamical stability is one of
essential criteria to select among multiple NE that are solutions to a game
and it is usually considered with respect to some adjustment process that
underlies the game. Selten \cite{selten} introduced the idea of
perturbations in normal form classical games to get rid of unreasonable NE
as solutions to these games by using a static procedure based on intuitive
equilibrium conditions of NE. Replicator dynamic \cite{cressman}\cite{taylor}
in evolutionary game theory is an adjustment process for which frequency of
a particular strategy evolves according to its payoff relative to that of
other strategies. For pure available classical strategies $S_{1},...,S_{n}$
let the proportion of players of $S_{i}$ at a particular time be $%
p_{i},(i=1,...,n),$ so that the average population strategy is the vector $%
\mathbf{p}=(p_{i})$, with the expected payoff of an $S_{i}$ player being $\
f_{i}(\mathbf{p)}$ and the overall expected payoff in the population being $%
F(\mathbf{p)=}\sum p_{i}f_{i}(\mathbf{p)}$. Then the replicator dynamic is
defined by the differential equation\cite{mark1}

\begin{equation}
\frac{dp_{i}}{dt}=p_{i}(f_{i}(\mathbf{p)-}F(\mathbf{p))}
\end{equation}
Thus the proportion of players which play the better strategies increase
with time. The replicator dynamic is originally based on biological
considerations but it has also found applications in Economics.
Evolutionarily Stable Strategies (ESSs) of evolutionary game theory are
attractors of replicator dynamic with an important property of stability.
When a population plays an ESS it cannot be invaded by a small number of
mutants.

Quantum games apart from finding applications in quantum information \cite
{werner} have also been considered interesting from another perspective.
Referring to Dawkins' 'Selfish Gene' \cite{dawkins} Eisert et al \cite
{eisert} hinted that games of survival are being played already on molecular
level where quantum mechanics dictates the rules. Self replication of DNA
molecules and synthesis of proteins which govern all the processes of a
living molecule are two essential ways \cite{patel} where quantum
information exchange can be said to play an important role. A consideration
of quantum games in such situations seems a good reason to make these games
interesting as indicated by Eisert et al \cite{eisert}. A decision by a DNA
molecule, for example to replicate itself or to synthesize a protein, can be
imagined as a set of instructions that can formulated as a quantum strategy.
How can, then, a DNA molecule evolves and develops a procedure to prepare a
particular protein, for example, and not some other; assuming the procedure
constitutes a quantum strategy. Why the other strategies, leading to
synthesis of a different protein for example, instead of the desired one are
not successful? This brings to mind the notion of mutant strategy from ESS
theory that tries to invade an ESS without success and ESS comes out as a
winning or successful strategy having a property of stability against
perturbations from mutants.

We assume life molecules in possession of many strategies out of which a few
or just one is successful in the sense of its increased use with time. The
notion of ESS derived from population biology problems but the concept
remains valid even when, for example, there are just two players to play a
bi-matrix game. In that case the ESS will be a strategy the players will
choose to play almost all the time and the fraction of the total time will
replace the usual term called `frequency' that corresponds to fraction of
the total population. The molecule deciding in favor of a particular set of
steps out of many available options gave us a good reason to consider
perturbations in quantum game like situation that we believe to exist in
that domain of intermolecular interactions. Mutant strategies constituting
perturbations in classical evolutionary game theory are not successful
against the Evolutionarily Stable Strategy (ESS) and ESS appears as a stable
strategy that cannot be invaded when mutants appear in small numbers.

Underlying process in the definition of ESS is replicator dynamic based on
Darwinian idea of `survival of the fittest' leading to increase with time
the frequency of a strategy giving better pay off. Replicator dynamic is a
simple and theoretically attractive process that can have, we believe, an
intuitive appeal to be extended to quantum game like situations. Among the
many possible quantum strategies that we suppose to be at the disposal of a
life molecule the winning strategy, corresponding to synthesis of desired
protein for example, can be imagined as a result of micro-level replicator
dynamic that becomes responsible for the increased use of the winning stable
strategy to such an extent that the molecule implements it almost all the
time and mutant set of procedures resulting in undesirable synthesis cannot
invade the stable solution called ESS coming out of replicator dynamic.

In this perspective the use of replicator dynamic to select among many NE a
special one corresponding to a stable solution, in quantum game like
situations that we believe to exist in molecules forming the basis of life,
appears an interesting hypothesis related to the mechanism existing in
molecules to formulate and evolve established and stable procedures to
perform their fundamental functions. An adjustment process having fruitful
results both in population biology problems and economics is simple and
intuitively appealing to form a mechanism to provide stable solutions in the
domain of molecular interactions.

We apply Marinatto and Weber's scheme \cite{marinatto} to a symmetric game
between two players. Our reason to select this scheme is that the idea of a
mixed NE and mixed ESS can easily be given a meaning without some extra
assumptions when the pure strategies of the quantized version of the game
are taken as the implementation by the players of identity $I$ and Pauli
spin-flip operator $\sigma _{x\text{ }}$respectively. In Marinatto and
Weber's scheme quantum `tactics' are applied by selecting $I$ and $\sigma
_{x}$ with classical probabilities similar to the case of mixed NE in
classical games where pure strategies are chosen with classical
probabilities. We then show how the `entanglement' usually considered as a
pure quantum phenomenon can be exploited to select and decide ESSs in
quantum version of a game satisfying a few requirements. The game we
consider is symmetric similar to corresponding examples in evolutionary game
theory where ESS is usually defined for symmetric pairwise contests.

In an earlier letter \cite{azhar} we showed that for certain asymmetric
quantum games between two players ESSs can be made to appear or disappear
via a control over the initial state even when the corresponding strategies
remain NE for those initial states. Our motivation in this letter is to
explore similar possibilities for symmetric quantum games. The ESS idea was
originally defined for symmetric games and will be of more interest when
entanglement and entangled states are exploited in those games as well. Even
for possible molecular level games referred to earlier the symmetric
contests seems more appropriate and interesting. It also brings the exciting
subject of mathematical theory of ESSs in the domain of quantum games and
also quantum mechanics in general. During the last thirty years the theory
of ESSs has been developed by collaboration between mathematicians and
evolutionary biologists. An extension of the ESS idea in quantum games is an
opportunity for both quantum game theory and classical theory of ESSs to
broaden their domains and also look for topics of common interest.

\section{Evolutionary Stability}

An Evolutionarily Stable Strategy (ESS) was originally defined by Smith and
Price \cite{smith} with the motivation that a population playing the ESS can
withstand a small invading group. They considered a symmetric game where the
players are anonymous. Let $P[x,y]$ be the payoff to a player playing $x$
against the player playing $y$.Strategy $x$ is an ESS if for any alternative
strategy $y$, the following two requirements are satisfied:

\begin{equation}
P[x,x]\geq P[y,x]  \label{ess1}
\end{equation}
and in the case were condition \ref{ess1} is satisfied as an equality:

\begin{equation}
P[x,y]>P[y,y]  \label{ess2}
\end{equation}

Requirement \ref{ess1} is in fact the Nash requirement and requires that no
single individual can gain by unilaterally changing strategy from $x$ to $y$%
. For a linear structure of the game the two requirements \ref{ess1} and \ref
{ess2} are equivalent to that for any $y\neq x$ there is a critical positive
value such that if the frequency of the $y$-mutant strategists is lower than
this value (while the rest of the population sticks to the strategy $x$), it
is better for everyone to stick to majority strategy $x$, better being
defined in terms of individual maximization of the payoff function $P$ \cite
{eshel}\cite{vickers}.Thus, as ESS is a strategy which, if played by almost
all members of a population, cannot be displaced by a small invading group
playing any alternative strategy. An ESS will persist as the dominant
strategy through time, so that strategies observed in the real world will
tend to be ESSs \cite{mark}.ESS comes out dynamically stable with respect to
replicator dynamic. In molecular level situations when replicator dynamic is
taken as the underlying process the ESSs involving quantum strategies will
similarly tend to be observed in real micro world.

\section{Evolutionary stability in symmetric quantum games}

In Marinatto and Weber's scheme \cite{marinatto} an entangled `initial
strategy' is forwarded to two players on which they apply their `tactics'.
To remain consistent with available literature on the theory of ESSs in
classical games we preferred to call Marinatto and Weber's `initial
strategy' an `initial state' and call their `tactics' as `strategies'. This
change in terminology has no effect on the originally proposed scheme to
play a quantum game \cite{personal}. We consider following symmetric
bi-matrix game between two players for two available classical strategies $%
S_{1}$ and $S_{2}$

\begin{equation}
\left( 
\begin{array}{cc}
(\alpha ,\alpha ) & (\beta ,\gamma ) \\ 
(\gamma ,\beta ) & (\sigma ,\sigma )
\end{array}
\right)  \label{matrix}
\end{equation}

with the constants of the matrix $\alpha ,\beta ,\gamma ,\sigma $ satisfying
the following conditions.

\begin{eqnarray}
\alpha ,\beta ,\gamma ,\sigma &\geq &0  \notag \\
(\sigma -\beta ) &>&0  \notag \\
(\gamma -\alpha ) &\geq &0  \notag \\
(\gamma -\alpha ) &<&(\sigma -\beta )  \label{conditions}
\end{eqnarray}

We take the initial quantum state to play the above game to be\cite
{marinatto}:

\begin{equation}
\left| \psi _{in}\right\rangle =a\left| S_{1}S_{1}\right\rangle +b\left|
S_{2}S_{2}\right\rangle
\end{equation}

where 
\begin{equation}
\left| a\right| ^{2}+\left| b\right| ^{2}=1
\end{equation}

The associated density matrix is\cite{marinatto}:

\begin{eqnarray}
\rho _{ini} &=&\left| a\right| ^{2}\left| S_{1}S_{1}\right\rangle
\left\langle S_{1}S_{1}\right| +ab^{\star }\left| S_{1}S_{1}\right\rangle
\left\langle S_{2}S_{2}\right|  \notag \\
&&+a^{\star }b\left| S_{2}S_{2}\right\rangle \left\langle S_{1}S_{1}\right|
+\left| b\right| ^{2}\left| S_{2}S_{2}\right\rangle \left\langle
S_{2}S_{2}\right|
\end{eqnarray}

Where $p$ and $q$ be the probabilities of two players to act with the
operator $\overset{\wedge }{I}$. Payoff to a $p$ player against a $q$ player
for the payoff matrix (\ref{matrix}) are written as \cite{marinatto}:

\begin{eqnarray}
P(p,q) &=&\alpha \left\{ pq\left| a\right| ^{2}+(1-p)(1-q)\left| b\right|
^{2}\right\} +  \notag \\
&&\beta \left\{ p(1-q)\left| a\right| ^{2}+q(1-p)\left| b\right|
^{2}\right\} +  \notag \\
&&\gamma \left\{ p(1-q)\left| b\right| ^{2}+q(1-p)\left| a\right|
^{2}\right\} +  \notag \\
&&\sigma \left\{ pq\left| b\right| ^{2}+(1-p)(1-q)\left| a\right|
^{2}\right\}
\end{eqnarray}

The condition that makes $(p^{\star },p^{\star })$ a NE is given as:

\begin{eqnarray}
&&P(p^{\star },p^{\star })-P(p,p^{\star })  \notag \\
&=&(p^{\star }-p)\left[ -\left| a\right| ^{2}(\sigma -\beta )+\left|
b\right| ^{2}(\gamma -\alpha )+p^{\star }\left\{ (\sigma -\beta )-(\gamma
-\alpha )\right\} \right] \geq 0  \notag \\
&&  \label{NE}
\end{eqnarray}

From the relation (\ref{NE}) it is clear that there can be three NE i.e. the
pure strategies $p^{\star }=0$, $p^{\star }=1$ and the mixed strategy $%
p^{\star }=\frac{(\sigma -\beta )\left| a\right| ^{2}-(\gamma -\alpha
)\left| b\right| ^{2}}{(\sigma -\beta )-(\gamma -\alpha )}$. In the earlier
form of this letter we only considered the strategy $p^{\star }=0$ and
showed that it is evolutionary stable in the classical game (\ref{matrix})
with conditions (\ref{conditions}) but not so in a quantum version of the
same game. We are grateful to anonymous referee who hinted the need to study
the problem in a more general and systematic way. We now consider the
evolutionary stability of these three NE of the symmetric game (\ref{matrix}%
) in three separate cases:

\subsection{Case $p^{\star }=0$}

For the strategy $p^{\star }=0$ to be a NE we should have:

\begin{equation}
P(0,0)-P(p,0)=\frac{p}{(\gamma -\alpha )+(\sigma -\beta )}\left[ \left|
a\right| ^{2}-\frac{(\gamma -\alpha )}{(\gamma -\alpha )+(\sigma -\beta )}%
\right] \geq 0  \label{differ}
\end{equation}

Now $[P(0,0)-P(p,0)]>0$ when $1\geq \left| a\right| ^{2}>\frac{(\gamma
-\alpha )}{(\gamma -\alpha )+(\sigma -\beta )}$ and $p^{\star }=0$ is a pure
ESS. However, at $\left| a\right| ^{2}=\frac{(\gamma -\alpha )}{(\gamma
-\alpha )+(\sigma -\beta )}$ we have the difference (\ref{differ}) zero and $%
p^{\star }=0$ can be an ESS if

\begin{eqnarray}
&&P(0,p)-P(p,p)  \notag \\
&=&p\left\{ (\gamma -\alpha )+(\sigma -\beta )\right\} \left\{ \left|
a\right| ^{2}-\frac{(1-p)(\gamma -\alpha )+p(\sigma -\beta )}{(\gamma
-\alpha )+(\sigma -\beta )}\right\} >0
\end{eqnarray}
it can be written as:

\begin{equation}
P(0,p)-P(p,p)=p\left\{ (\gamma -\alpha )+(\sigma -\beta )\right\} \left\{
\left| a\right| ^{2}-\digamma \right\} >0
\end{equation}

Where $\frac{(\gamma -\alpha )}{(\gamma -\alpha )+(\sigma -\beta )}\leq
\digamma \leq \frac{(\sigma -\beta )}{(\gamma -\alpha )+(\sigma -\beta )}$
for the range $0\leq p\leq 1.$ In such a situation $p^{\star }=0$ can be an
ESS only when $\left| a\right| ^{2}>\frac{(\sigma -\beta )}{(\gamma -\alpha
)+(\sigma -\beta )}$ which is not possible because $\left| a\right| ^{2}$ is
fixed at $\frac{(\gamma -\alpha )}{(\gamma -\alpha )+(\sigma -\beta )}.$
Therefore $p^{\star }=0$ is a stable NE or an ESS for $1\geq \left| a\right|
^{2}>\frac{(\gamma -\alpha )}{(\gamma -\alpha )+(\sigma -\beta )}$ and at $%
\left| a\right| ^{2}=\frac{(\gamma -\alpha )}{(\gamma -\alpha )+(\sigma
-\beta )}$ this NE becomes unstable. The classical game is obtained by
fixing $\left| a\right| ^{2}=1$ for which $p^{\star }=0$ is a stable NE.
However, this NE does no remain stable when $\left| a\right| ^{2}=\frac{%
(\gamma -\alpha )}{(\gamma -\alpha )+(\sigma -\beta )}$ corresponding to an
entangled initial state; though, the NE remains intact in both forms of the
game.

\subsection{Case $p^{\star }=1$}

Similar to previous case we write the NE condition for the strategy $%
p^{\star }=1$ as:

\begin{equation}
P(1,1)-P(p,1)=\frac{(1-p)}{(\gamma -\alpha )+(\sigma -\beta )}\left[ -\left|
a\right| ^{2}+\frac{(\sigma -\beta )}{(\gamma -\alpha )+(\sigma -\beta )}%
\right] \geq 0  \label{diff2}
\end{equation}

Now $p^{\star }=1$ is a pure ESS for $0\leq \left| a\right| ^{2}<\frac{%
(\sigma -\beta )}{(\gamma -\alpha )+(\sigma -\beta )}.$ At $\left| a\right|
^{2}=\frac{(\sigma -\beta )}{(\gamma -\alpha )+(\sigma -\beta )}$ we have
the difference given in eq. (\ref{diff2}) zero. The strategy $p^{\star }=1$,
then, becomes an ESS when

\begin{eqnarray}
&&P(1,p)-P(p,p)  \notag \\
&=&(1-p)\left\{ (\gamma -\alpha )+(\sigma -\beta )\right\} \left\{ -\left|
a\right| ^{2}+\frac{(1-p)(\gamma -\alpha )+p(\sigma -\beta )}{(\gamma
-\alpha )+(\sigma -\beta )}\right\} >0  \notag \\
&&
\end{eqnarray}

It is possible only when $\left| a\right| ^{2}<\frac{(\gamma -\alpha )}{%
(\gamma -\alpha )+(\sigma -\beta )}.$ Therefore the strategy $p^{\star }=1$
is a stable NE for $0\leq \left| a\right| ^{2}<\frac{(\sigma -\beta )}{%
(\gamma -\alpha )+(\sigma -\beta )}.$ Therefore, the strategy $p^{\star }=1$
not stable classically ,i.e. for $\left| a\right| ^{2}=1$, can be stable for
an entangled initial state.

\subsection{Case $p^{\star }=\frac{(\protect\sigma -\protect\beta )\left|
a\right| ^{2}-(\protect\gamma -\protect\alpha )\left| b\right| ^{2}}{(%
\protect\sigma -\protect\beta )-(\protect\gamma -\protect\alpha )}$}

In case of the mixed strategy

\begin{equation}
p^{\star }=\frac{(\sigma -\beta )\left| a\right| ^{2}-(\gamma -\alpha
)\left| b\right| ^{2}}{(\sigma -\beta )-(\gamma -\alpha )}  \label{mixed}
\end{equation}

we have from the NE condition (\ref{NE})

\begin{equation}
P(p^{\star },p^{\star })-P(p,p^{\star })=0  \label{balance}
\end{equation}

The mixed strategy (\ref{mixed}) becomes an ESS when

\begin{eqnarray}
&&P(p^{\star },p)-P(p,p)  \notag \\
&=&(p^{\star }-p)\left[ -\left| a\right| ^{2}(\sigma -\beta )+\left|
b\right| ^{2}(\gamma -\alpha )+p\left\{ (\sigma -\beta )-(\gamma -\alpha
)\right\} \right] >0  \label{ess2N}
\end{eqnarray}

for all $p\neq p^{\star }$. Write now the strategy $p$ as $p=p^{\star
}+\bigtriangleup $. For the mixed strategy of eq. (\ref{mixed}) the payoff
difference of the eq. (\ref{ess2N}) can be reduced to:

\begin{equation}
P(p^{\star },p)-P(p,p)=-\bigtriangleup ^{2}\left\{ (\sigma -\beta )-(\gamma
-\alpha )\right\}
\end{equation}

So that, for the game defined in the conditions (\ref{conditions}) the mixed
strategy $p^{\star }=\frac{(\sigma -\beta )\left| a\right| ^{2}-(\gamma
-\alpha )\left| b\right| ^{2}}{(\sigma -\beta )-(\gamma -\alpha )}$ cannot
be an ESS, though it can be a NE of the symmetric game.

\section{Conclusion}

Dynamic stability is a well known refinement concept in situations where
multiple NE arise as solutions to a symmetric quantum game played between
two players. ESS is a stable NE in a game when its dynamics puts a direct
proportionality between frequency of a strategy and its relative payoff
advantage usually termed as replicator dynamic. We suggested that this idea
can be interesting not only in population biology but also in quantum game
theory. We explored evolutionary stability of NE in a symmetric quantum game
played between two players via a scheme proposed recently by Marinatto and
Weber \cite{marinatto}. We showed that in this scheme the evolutionary
stability of a pure strategy for a symmetric game can be changed by
maneuvering the initial state even when the strategy remains a NE during
such a maneuver. For example, in a symmetric game a pure strategy can be an
ESS in the classical form of the game but it does not remain so in some
quantized form of the same game. We also showed that in the proposed scheme 
\cite{marinatto} with its particular form of initial state the evolutionary
stability of a mixed strategy, however, cannot be changed while retaining
the corresponding NE. Our results show that entangled states have also
potential roles to play in the stability properties of NE in symmetric
games. It shows the possibility of achieving dynamic stability of NE in
symmetric quantum games via the use of entangled states.

\section{Acknowledgment}

One of us (A.I) is grateful to Pakistan Institute of Lasers and Optics,
Islamabad, for support.


\begin{thebibliography}{99}
\bibitem{eisert}  Jens Eisert, Martin Wilkins, and Maciej Lewenstein,
Quantum Games and Quantum Strategies. Phy. Rev. Lett. 83, 3077. Also
quant-ph/0004076

\bibitem{marinatto}  L. Marinatto and T. Weber, Phys. Lett. A 272, 291 (2000)

\bibitem{selten}  Selten, R. (1975). Reexamination of the Perfectness
Concept for Equilibrium Points in Extensive Games. Int. J. Game Theory 4,
25-55.

\bibitem{werner}  R.F.Werner, Phy. Rev, A 58, 1827 (1998)

\bibitem{dawkins}  R. Dawkins. The Selfish Gene. (Oxford University Press,
Oxford, 1976)

\bibitem{smith}  Maynard Smith, J. and Price, G.R. (1973) The Logic of
Animal Conflict. Nature, 246, 15-18

\bibitem{eshel}  I, Eshel. On the Changing Concept of Evolutionary
Population Stability as a Reflection of a Changing Point of View in the
Quantitative Theory of Evolution. J. Math. Biol. (1996) 34: 485-510

\bibitem{vickers}  G.T. Vickers and C. Canning. On the Definition of an
Evolutionarily Stable Strategy. J. theor. Biol. (1987) 129, 349-353.

\bibitem{mark}  M. Broom. Bounds on the Number of ESSs of a Matrix Game.
Math. Biosci., 167 (2000) 163-175.

\bibitem{personal}  Marinatto, L. Personal correspondence.

\bibitem{cressman}  R. Cressman and K.H.Schlag, Dynamic Stability in
Perturbed Games. Discussion paper No. B-321. Rheinsche
Friedrich-Wilhelms-Universitat Bonn.

\bibitem{patel}  A. Patel, Quantum Algorithms and the Genetic Code.
quant-ph/0002037

\bibitem{taylor}  Taylor, P.D., Jonker, L.B., 1978. Evolutionarily Stable
Strategies and Game Dynamics. Math. Biosci. 40, 145-156

\bibitem{mark1}  Broom, M. , Canning, C. \& Vickers, G.T. (1997).
Muti-player matrix games. Bull. Math. Biol. 59, 931-952

\bibitem{azhar}  A. Iqbal and A.H.Toor, Evolutionarily Stable Strategies in
Quantum Games. Physics Letters A 280 (2001) 249. quant-ph/0007100
\end{thebibliography}
\end{document}